\documentclass[twoside,reqno]{bjp}
\usepackage{epsfig,cite}
\usepackage{graphics}
\usepackage{amssymb,amsmath}
\usepackage{times}
\usepackage{mathrsfs}
\usepackage{color}
\begin{document}
	
\sloppy \raggedbottom
\setcounter{page}{1}

\title{From Black Holes to Baby Universes: Exploring the Possibility\\
of Creating a Cosmos in the Laboratory}

\runningheads{From Black Holes to Baby Universes}{S.~Ansoldi, Z.~Merali, E.~I.~Guendelman}
	
\begin{start}
\author{S.~Ansoldi\thanks{ansoldi@fulbrightmail.org}}{1,2,3}

\coauthor{Z.~Merali\thanks{merali@fqxi.org}}{4}

\coauthor{E.~I.~Guendelman\thanks{guendel@bgu.ac.il}}{5,6,7}

\address{Istituto Nazionale di Fisica Nucleare (INFN), Sezione di Trieste, 34149 Padriciano (TS), Italy}{1}

\address{International Center for Relativistic Astrophysics (ICRA), Piazza della Repubblica 10, 65122 Pescara (PE), Italy}{2}

\address{Universit\`{a} degli Studi di Udine, Dipartimento di Scienze Matematiche Informatiche e Fisiche, 33100 Udine (UD), Italy}{3}

\address{Foundational Questions Institute (FQXi), PO Box 3055, Decatur, GA 30031, USA}{4}

\address{Ben Gurion University of the Negev, Department of Physics, Beer-Sheva, Israel}{5}

\address{Bahamas Advanced Study Institute and Conferences, 4A Ocean Heights, Hill View Circle, Stella Maris, Long Island, The Bahamas}{6}

\address{Frankfurt Institute for Advanced Studies, Giersch Science Center, Campus Riedberg, Frankfurt am Main, Germany}{7}

\received{\today}

\begin{Abstract}
We discuss the essential features of baby-universe production, starting from a description of black holes and wormholes, in terms of the causal structure of spacetime, and following a qualitative review of the connection between vacuum decay and inflation in early universe cosmology. Related open questions are also addressed, especially in connection with the possibility that baby universes could be formed within our present universe -- either in a future particle accelerator, or spontaneously.
\end{Abstract}
\end{start}

\section{Introduction}

The word cosmology stems from the Greek words $\kappa{}\acute{o}{}\sigma{}\!\mu{}o\zeta$ (meaning ``order'', as opposed to chaos, or ``world'') and $\lambda{}\acute{o}{}\gamma{}o{}\zeta$ (which means ``word'', or ``discourse''), and cosmology is thus the ancient
study of the universe as an ordered whole. The discipline was revolutionised in the twentieth century with the advent of Einstein's general theory of relativity, which had a dramatic impact on our understanding of the evolution of the universe. Where previously it was assumed that matter and energy interact against the static backdrop of space and time, it has now became apparent that space and time are intertwined, and that the expansion of the universe does not take place in ``spacetime'', it actually \emph{creates} spacetime.

From this point of view, it is perhaps unsurprising that fundamentally new processes can take place in the context of general relativity. We will, here, analyze the physics of black holes and wormholes, and their relationship to the proposed generation of ``baby universes'' (also referred to in the literature as ``child universes'') --- regions of spacetime that are initially connected to our own universe, but which causally disconnect from us and inflate, becoming self-contained cosmoses.  In the 1980s, it was proposed that such baby universes could, in theory, be manufactured in a future particle accelerator and that they would appear from the outside to be miniature black holes~\cite{bib:2017BigBanLitRooMer}. More recently, it has been suggested that baby universes may form spontaneously in our universe and that, if this process occurs, it may have implications for the development of a theory of quantum gravity, unifying general relativity with the physics of the micro-realm, quantum mechanics~(see, e.g.,~\cite{bib:1993NucPhy__B305208Amb}).

In what follows, we will first discuss how black holes are tightly connected with the causal structure of spacetime (section~\ref{sec:bhw}). We will introduce the concepts of the chronological or causal past and future of a subset of events in spacetime, and, then, focus our attention on the basic elements that can be used to describe the asymptotic structure of spacetime, i.e. the spacetime structure at infinity (both, in space and time). These concepts are central to a rigorous definition of a black hole; moreover, they can be naturally related to an intuitive understanding of these objects. We shall also define the related concept of a wormhole -- a tunnel linking two regions of spacetime -- which plays an integral role in baby-universe formation.

In section~\ref{sec:inf}, we will outline the motivation for the theory of inflation, which forms the current cosmological paradigm, and also provides the mechanism for baby-universe generation. We will then explain how inflation theory gives rise to the baby-universe proposal, in section~\ref{sec:buf}. (This process is also known, in the literature, as spacetime tunneling with black hole/wormhole creation.) We will provide an essential account of the most relevant physical aspects of the process, with an emphasis on the relationship with the physics of the vacuum.

Section~\ref{sec:con} will, finally, contain a short summary of the main ideas discussed in this contribution, and a qualitative description of some open problems in the field, including: the feasibility of making a baby universe in a real-world particle accelerator; proposed signatures of lab-made universes; and the possibility of baby universes arising spontaneously in the vacuum (and the implications of such a process for quantum gravity).


\section{\label{sec:bhw}Black Holes and Wormholes}

The conception of the idea of a black hole -- as a region from which
not even light can escape -- perhaps surprisingly, predates general relativity. In 1784, the English clergyman and natural philosopher John Michell, considering Newton's corpuscular theory of light,
pondered whether `light particles' may interact with gravity and would thus have an associated escape velocity. (The escape velocity is the speed that is sufficient for a body to be able to move arbitrarily far away from another body, that, for simplicity, we can consider much more massive. For instance, the escape velocity from the Earth is about $11\;\mathrm{km}/\mathrm{sec}$, while the escape velocity from the Sun is about $620\;\mathrm{km}/\mathrm{sec}$.)

Michell calculated that
\begin{quote}
	``\dots if the semi-diameter of a sph\ae{}re of the same density with the sun were to exceed that of the sun in the proportion of $500$ to $1$, a body falling from an infinite height towards it, would have acquired at its surface a greater velocity than that of light, and consequently, supposing light to be attracted by the same force in proportion to its vis inerti\ae{}, with other bodies, all light emitted from such a body would be made to return towards it, by its own proper gravity \dots''~\cite{bib:1784PhiTraRoySoc_74_35Mic}
\end{quote}

Pierre-Simon Laplace also independently developed a notion of Newtonian black holes, or ``dark stars''~\cite{bib:1799AllGeoEph__4__1Lap}. Note that, in this historical context, there was no modern conception of spacetime, only Newtonian space plus time. There was also no bound to the speed of propagation of interactions: action at a distance allows for the physical effects of perturbations to propagate instantaneously across all space. The idea of a massive body that can trap light is thus not tied in any way to a dynamical spacetime.

This situation, of course, changes drastically with the general theory of relativity; although it took about half a century after the theory was proposed for the discovery of what we, nowadays, consider the first mathematically rigorous realization of a black hole (the Schwarzschild solution)~\cite{bib:1916SitKonPreAkaWisBer189}, to provide a clear understanding of spacetime structure related to black holes. The reason for this long process (please, see~\cite{bib:1987ThrHunYeaGra199Isr} for a detailed account of it), can be substantially traced back to the fact that it is standard, in pre-relativistic physics, to associate a metric interpretation to the values of coordinates, and associate, in this way, a direct physical meaning to the values of tensor components. This approach is, however, very dangerous (and conceptually wrong) in general relativity. The gradual recognition of this danger went together with the development of different coordinate systems that progressively clarified distinctive properties of the Schwarzschild solution; this fact is also witnessed by the change of the name of these objects to ``black holes''.

\begin{figure}
	\begin{center}
	\includegraphics[width=9cm]{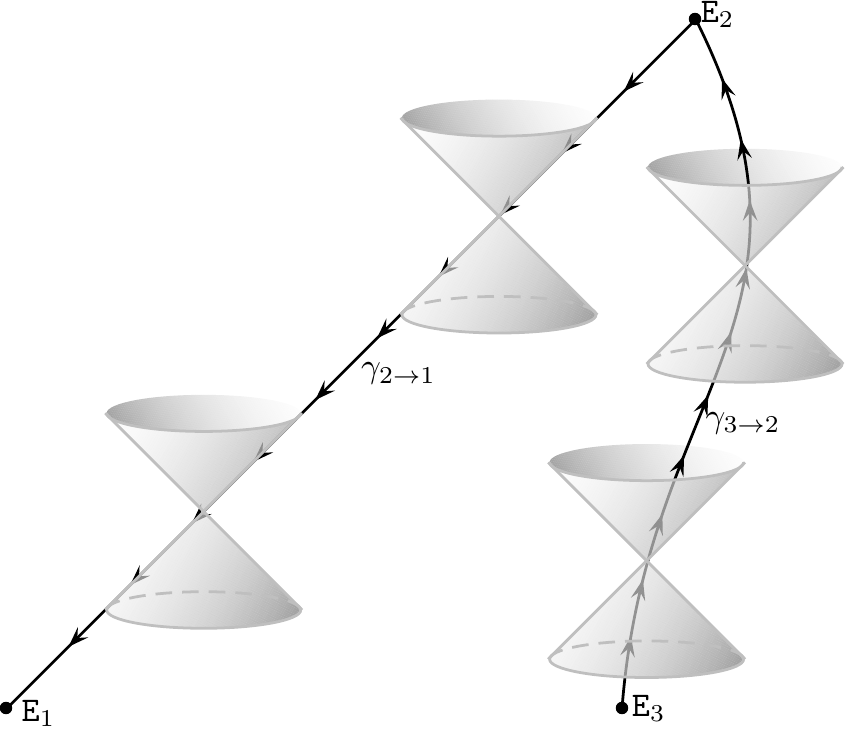}
	\caption{\label{fig:crocaupasfut}In this figure we exemplify some of the concepts described in the main text. For graphical convenience, we chose a coordinate system, in which the light cones have the same shape as in special relativity, when the speed of light is set to $1$. The points \texttt{E}$_{1}$, \texttt{E}$_{2}$, and \texttt{E}$_{3}$ are events in spacetime. The curve $\gamma _{3 \to 2}$ is an everywhere timelike curve (at every point the vector tangent to the curve falls inside the light cone). The arrows show the \emph{past}$\;\to\;$\emph{future} direction, i.e., this curve is future directed. Since, it connects the event \texttt{E}$_{3}$ with the event \texttt{E}$_{2}$, according to the definition, \texttt{E}$_{2}$ is in the chronological future of \texttt{E}$_{3}$. On the curve $\gamma _{2 \to 1}$ the arrows point, instead, in the \emph{future}$\;\to\;$\emph{past} direction. The curve is a causal curve: in particular it is nowhere timelike, and everywhere lightlike (at every point the tangent vector to the curve falls exactly on the light cone). According to the definitions given in the main text, \texttt{E}$_{1}$ is in the causal past of \texttt{E}$_{2}$. Intuitively, we can understand that there can be no past directed timelike curves connecting \texttt{E}$_{2}$ to \texttt{E}$_{1}$: thus \texttt{E}$_{1}$ is not in the chronological past of \texttt{E}$_{2}$.}
	\end{center}
\end{figure}

In what follows, we will qualitatively discuss an intuitive definition of a black hole, in terms of the causal structure of spacetime, following the approach that can be found in the classical book by Hawking and Ellis~\cite{bib:1975CamUniPreHawEll}. Several ingredients are central in this understanding, and will be introduced below. One of them is the identification of events in a given spacetime ${\mathcal{M}}$ that can be affected by another event, or by a subset of them. This is what goes under the name of the \emph{chronological} (resp., \emph{causal}) \emph{future} of a set ${\mathcal{U}} \subset {\mathcal{M}}$ of events: this is the set of all events that can be connected by a \emph{future directed timelike} (resp., \emph{non spacelike}) curve starting from a given event in the set. Thus, if an event \texttt{q} can be reached, starting from a given event \texttt{p}, by moving along a future directed curve that at every point remains inside (resp., not outside) the light cone, then we say that \texttt{q} is in the chronological (resp., causal) future of \texttt{p}. Of course, analogous characterizations can be given for the \emph{chronological} (resp., \emph{causal}) \emph{past}, and we will indicate the causal past of the set ${\mathcal{U}} \subset {\mathcal{M}}$ by $J ^{-} ( {\mathcal{U}} )$.

A second ingredient is the understanding that, given an event \texttt{p} in spacetime, it could be possible to find one (or more) subsets ${\mathcal{A}} \subset {\mathcal{M}}$ of events to the past of \texttt{p}, such that the knowledge of what happens at these events is enough to completely determine what happens at \texttt{p}. In this case, the event \texttt{p} is in what is called the \emph{future domain of dependence of} ${\mathcal{A}}$. Again, a corresponding definition can be given for the \emph{past domain of dependence} of \texttt{p}, or, in general, of a larger set of events. It is thanks to concepts like the ones briefly outlined above, that we are able to rigorously speak of the causal relationships among events in spacetime.

\begin{figure}
	\begin{center}
		\includegraphics[height=6.0cm]{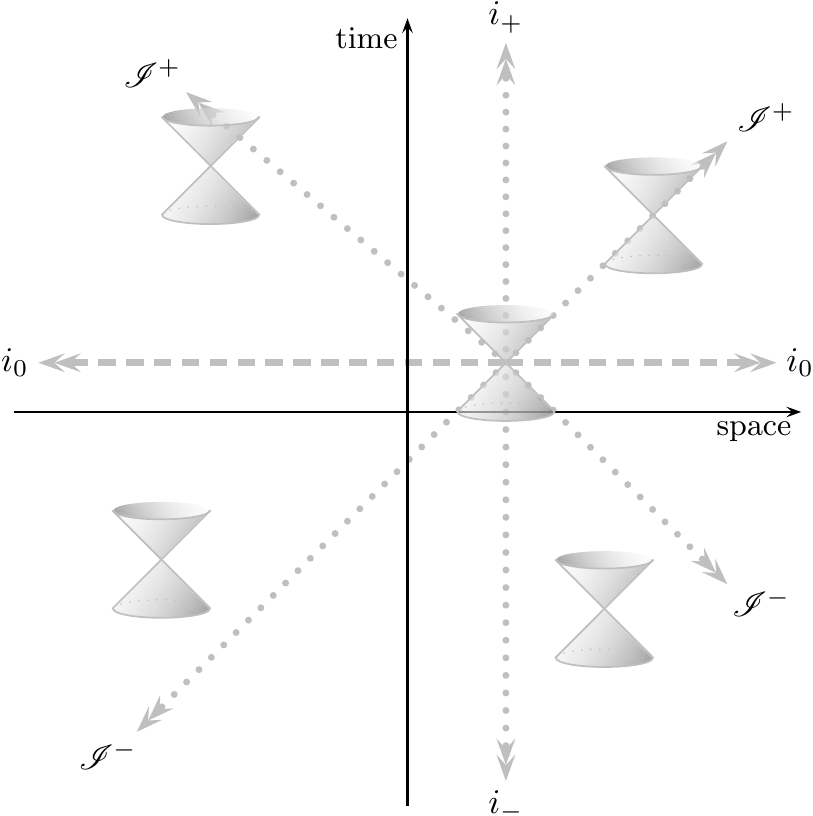}
		\hskip 1.9cm
		\includegraphics[height=6.0cm]{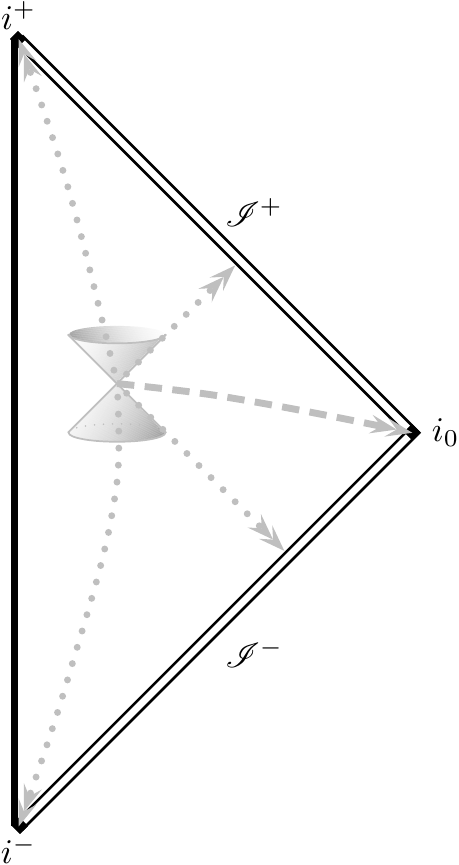}
		\caption{\label{fig:infstr}On the left we show the various types of infinity in Minkowski spacetime. The light cone structure identifies timelike, spacelike, and null vectors at every event. Starting from every event, we can remain at the same spatial position and move (dotted lines) to infinity in time toward the future (reaching $i ^{+}$, timelike future infinity) or toward the past (timelike past infinity, $i ^{-}$). Or, we can consider, at fixed time, points far away in space, reaching spacelike infinity, $i _{0}$. Another possibility is to move at infinity, both, in space and time, at the speed of light: toward the future we, then, reach future null infinity, ${\mathscr{I}} ^{+}$, while toward the past we reach past null infinity, ${\mathscr{I}} ^{+}$. It is common to perform a suitable (conformal) transportation to obtain a compact version (Penrose diagram) of (half of) the diagram shown on the left. This is shown on the right, where the notation is the same used on the left; in addition, the fact that spacetime is empty/flat at infinity is explicitly indicated by the double diagonal lines.}
	\end{center}
\end{figure}

There is, however, another fundamental aspect in our intuitive understanding of a black hole, and it is the fact that, in some sense, it has the possibility to trap us. To make this concept rigorous is more subtle, because, conceptually, a \emph{test} body that can ``only'' move several light years away from a given point in space (given enough time, of course) is as trapped to this point as a body that can just move a few millimeters away. This is why the concept of being trapped inside some region is more conveniently expressed as the \emph{impossibility to move as far away as we want}, or, in other words, to move \emph{infinitely far away}. We thus realize that, at least at the very first stage, it might be necessary to consider the asymptotic structure of spacetime to be able to properly understand the concept of a trapping region as a global property of spacetime. A first, convenient, assumption, is to imagine spacetimes that have a structure at large distances similar to the one of Minkowski spacetime, which we are familiar with.~\footnote{It is possible to generalize these ideas, but we do not really need such generalizations here.} We, then, come back to the unified understanding of spacetime given to us by special relativity. Indeed, several ways to go to \emph{infinity} exist in Minkowski spacetime. We can consider events as far away as possible from a given point, while keeping the time fixed: this boundary of spacetime is called \emph{spacelike infinity}, $i _{0}$. Or, we could wait an infinite time, while sitting at a fixed point in space: this is also a boundary of spacetime, which is called \emph{future timelike infinity}, $i ^{+}$.~\footnote{A corresponding definition can be given, of course, for \emph{past timelike infinity.}, $i ^{-}$.} And, finally, we could keep moving away from a given event at the speed of light, reaching what is called \emph{future null infinity}, ${\mathscr{I}} ^{+}${}.~\footnote{Again \emph{past null infinity}, ${\mathscr{I}} ^{-}$ is defined in a similar way.}

Now, let us assume that we can describe the whole spacetime ${\mathcal{M}}$ as a continuous sequence of spacelike hypersurfaces ${\mathcal{S}} (\tau)$, indexed by the (continuous, real) parameter $\tau$: ${\mathcal{M}} = \cup _{\tau} {\mathcal{S}} (\tau)$. This description does not need to be unique (and, indeed, it is \emph{not}!), so we can just pick one of the many that are possible. The parameter $\tau$ can be interpreted as time, and ${\mathcal{S}} (\tau)$ as a constant-time (hyper-)surface. Let us also assume that the spacetime ${\mathcal{M}}$ has one or more regions with the asymptotic structure defined above. Let us now consider one ${\mathcal{S}} (\tau)$. We, then, consider the set of all events that are in the causal past of future null infinity, which according to our previous notation is written as $J ^{-} ({\mathscr{I}} ^{+})$. We call ${\mathcal{B}} (\tau)$ what remains after removing this last set of events from ${\mathcal{S}} (\tau)$:
\[
	{\mathcal{B}} (\tau) = {\mathcal{S}} (\tau) - J ^{-} ({\mathscr{I}} ^{+})
	\ .
\]
By definition, an event \texttt{b} inside ${\mathcal{B}} (\tau)$ cannot be connected by any causal curve to ${\mathscr{I}} ^{+}$: if such a curve would exist, \texttt{b} would be inside $J ^{-} ({\mathscr{I}} ^{+})$, but we have obtained ${\mathcal{B}} (\tau)$ exactly by removing such events from ${\mathcal{S}} (\tau)$. If it is not empty, ${\mathcal{B}} (\tau)$ realizes, at time $\tau$, a subset of events that cannot send any test signal to future null infinity, and (a connected component in) it is called a \emph{black hole} on ${\mathcal{S}} (\tau)$: it certainly realizes our original intuition! Then, the \emph{event horizon} can be defined as the boundary of $J ^{-} ({\mathscr{I}} ^{+})$, i.e. a set of points that can just barely send a light signal to ${\mathscr{I}} ^{+}$.

Of course, the sets ${\mathcal{B}} (\tau)$ and the boundary of $J ^{-} ({\mathscr{I}} ^{+})$ could be empty sets. However, the Schwarzschild solution is a realization of a spacetime, in which they are not, and it represents the earliest example of a black hole spacetime. We also notice that the Schwarzschild solution contains two asymptotic regions that have a structure similar to the structure at the infinities of Minkowski spacetime.~\footnote{In general, spacetimes with more than two such regions can also exist, and there are also spacetimes with an infinity of such regions.}

\begin{figure}
	\begin{center}
		\includegraphics[width=11cm]{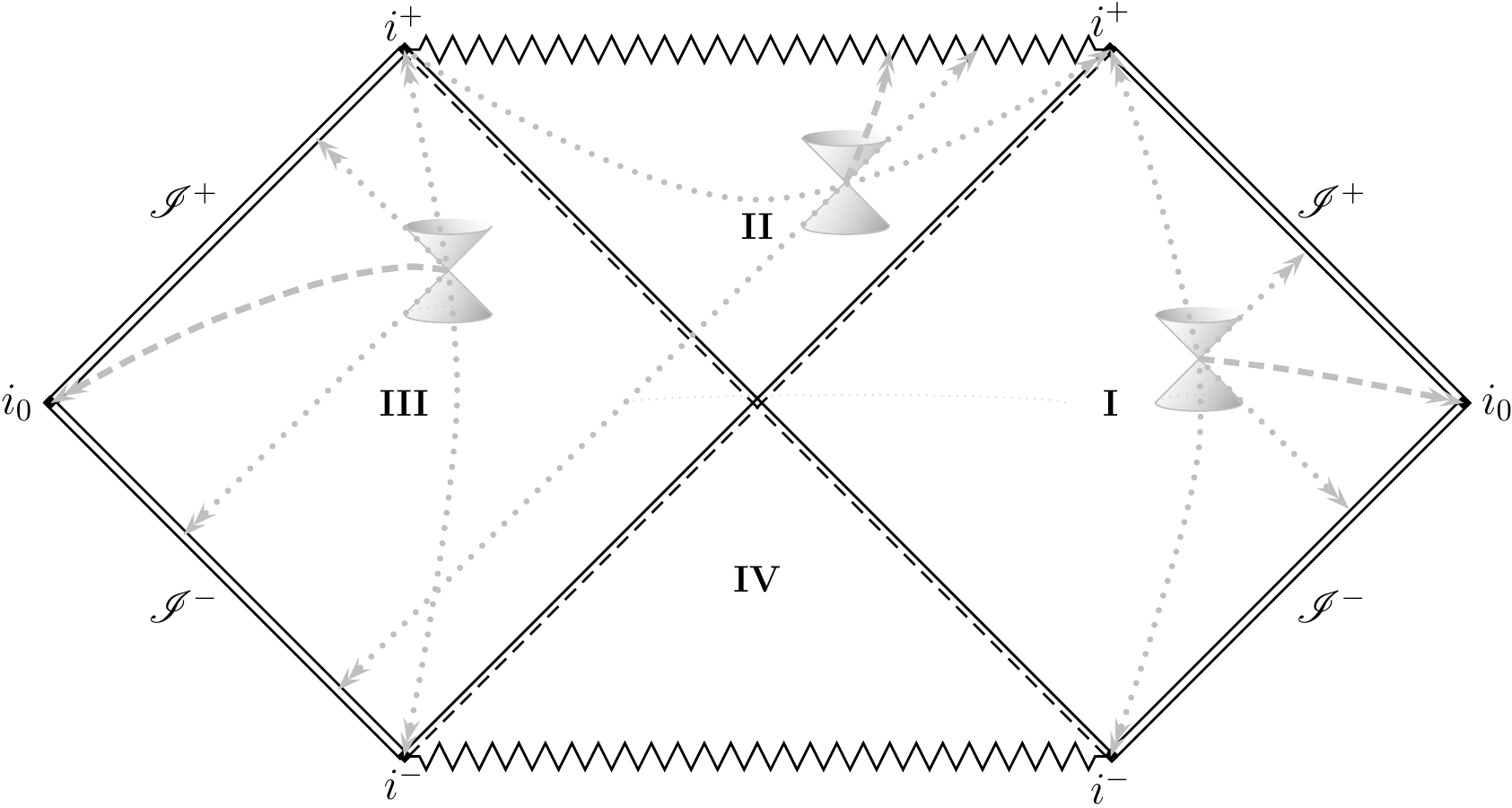}
		\caption{\label{fig:Schsol}Representation of the Schwarzschild solution in terms of a Penrose diagram. The notation is the same as in figure~\ref{fig:infstr}. We first note that the Schwarzschild spacetime has \emph{two} asymptotic regions (labels I and II in the figure) with properties similar to the asymptotic structure of Minkowski spacetime (cf. the Penrose diagram in figure~\ref{fig:infstr}). These regions are causally disconnected, as there is no causal curve starting inside in one of them, and ending in the other. We now focus on region I, which is called the \emph{black hole} region. If we consider the causal past of the future null infinity in region I, $J ^{-} ({\mathscr{I}} ^{+})$, we see that it consists of regions I and IV. Region II is thus not in $J ^{-} ({\mathscr{I}} ^{+})$. This is also true if we consider the causal past of the future null infinity in region III. Moreover, for every event in II it is true that even the fastest light signal will reach the \emph{zig-zagged} line on the top, which is the Schwarzschild singularity. For some events, the corresponding light cones are drawn together with curves representing events with the same time of the event at the center of the light cone but at a different place (dashed lines), events at the same place but at different times (dotted curves), and light rays directed toward the past and the future (dotted lines). Note that in regions I and III (the static regions), curves of constant time are spacelike (i.e., they flow in the left$\;\leftrightarrow\;$right direction), while curves of constant position are timelike (i.e., they flow in the bottom$\;\to\;$top direction). However, inside the black hole region, the opposite occurs, and the singularity is spacelike. The presence of two causally disconnected regions with the same asymptotic structure of Minkowski (I and III in the picture) is at the heart of the idea of wormholes (see the main text for details).}
	\end{center}
\end{figure}
This general definition of a black hole, in terms of causal structure, has a concrete realization in solutions to the Einstein equations that can be written in closed form. The first of these solutions was found by Karl Schwarzschild as early as 1915, and it is the first, and typical, example of a black hole. If we consider the form of the metric in the coordinates $(t,r,\theta,\varphi)$ (Schwarzschild coordinates,~\footnote{These coordinates are adapted to the symmetries that characterize Schwarzschild spacetime, which is static (outside the event horizon), and spherically symmetric (consistently, the meaning of the $\theta$ and $\varphi$ angles is the same as for polar coordinates in flat spacetime). The \emph{radial coordinate}, $r$, is called the \emph{circumferential radius}, because circles with constant $r$ have a length equal to $2 \pi r$; note, however, that the distance between one point on such a circle, and the center, is \emph{not} $r$. In a curved spacetime, it is not possible, in general, to find a radial coordinate that satisfies both the properties considered above, and this shows how important it is to consider the precise definition of coordinates, without relying on their names only.} $G$ is Newton's gravitational constant and $c$ is the speed of light)
\[
	d s ^{2}
	=
	-
	\left( 1 - \frac{2 G M}{c ^{2} r} \right) d t ^{2}
	+
	\left( 1 - \frac{2 G M}{c ^{2} r} \right) ^{-1} d r ^{2}
	+
	r ^{2} \left( d \theta ^{2} + \sin ^{2} \theta d \varphi ^{2} \right)
	,
\]
we recognize that the above expression has two problematic regions, corresponding to the values $r = 0$ and $r = 2 G M / c ^{2}$. Closer and closer to $r = 0$ the spacetime curvature takes larger and larger values, and this point is a physical singularity of spacetime. On the other hand $r = 2 G M / c ^{2}$, the gravitational radius, is not such a critical point. For instance, if we change coordinates, we can write the metric in other forms, where the problem at the gravitational radius disappears; however, the one at $r = 0$ always remains. All points identified by a radial coordinate equal to the gravitational radius form the ``event horizon'' of the Schwarzschild black hole. This region behaves as a one-way membrane, that can be crossed when falling toward the singularity at $r = 0$, but that cannot be crossed in the opposite direction. After crossing the event horizon moving to decreasing values of the radial coordinate $r$, the observer cannot reach (nor send any signal) to the outside anymore, and he/she is forced to keep moving toward $r = 0$.

In our universe, black holes are realized as stellar mass and supermassive black holes.
Stellar-mass black holes are formed by the gravitational collapse of stars whose masses exceed a certain limit. When a star reaches the end of its life and is no longer able to stop the inward pull of gravity using the outward pressure generated by nuclear fusion, different objects can be formed. White dwarfs, for example, are stellar remnants formed from the gravitational collapse of lower mass stars,
in which the outward pressure of a degenerate gas of electrons counteracts the pull of gravity. For a non-rotating object, the maximum mass is given by the Chandrasekhar limit, which is about $1.44$ solar masses.
For bigger masses of the progenitors, however, the pull of gravity is too strong and gravitational collapse continues. It could possibly be stopped when the density reaches nuclear densities; for masses up to a few solar masses, a neutron star is then formed. For even more massive objects, it is expected that gravitational collapse cannot be stopped, and a stellar mass black hole is then formed.

Black holes have been proposed to exist in a wide range of masses, from micro black holes that could have been formed as primordial black holes during the early stages of evolution of the universe~\cite{bib:1971MonNotRoyAstSoc152_75Haw,bib:1981ProThePhy_651443Sat} (or, indeed, may be generated in a particle accelerator due to exotic physics effects~\cite{bib:2017BigBanLitRooMer})
to supermassive black holes of billions of solar masses, such as the ones expected to be present at the center of galaxies. 

Another exciting possibility for the spacetime structure, which is suggested by the presence of more than one asymptotic region, are \emph{wormholes}. Technically, wormholes are allowed by the fact that the Einstein equations, locally, fix the geometry of spacetime, but they do not fix spacetime topology. Following Fuller and Wheeler~\cite{bib:1962PhyRev128919Ful}, without changing the geometry of spacetime, it is then possible to imagine the two, identical, asymptotic regions that we considered above, as joined by a throat, also called an \emph{Einstein-Rosen bridge}, or a \emph{Schwarzschild wormhole}: we obtain, in this way, a multiply connected spacetime. After this seminal work, the idea of a wormhole was revived about a quarter of a century later by Morris and Thorne~\cite{bib:1988AmeJouPhy_56395Mor}, who put forward the idea of traversable wormholes. The physics of wormholes then flourished, encouraging both the consideration of conceptually challenging realizations (such as, time machines) and technical analysis (especially related to their stability).
This will turn out to be a key ingredient for the idea of a baby universe, as we will discuss in the following section~\ref{sec:buf}.

Although wormholes are also allowed by general relativity, as opposed to black holes there is no evidence that they exist. It has been noted that for a wormhole to be traversable -- that is, allowing the passage of particles (or observers) from one region of spacetime to another -- they would need to be propped open and stabilized by exotic matter, with negative energy~\cite{bib:1989PhyRev__D_39318Vis,bib:1999PhyLet__B260175Ida}. It has also been proposed that (some) black holes may serve as gateways to other universes~\cite{bib:2016JouCosAstPhy160060Gar,bib:1999TheLifCosSmo}, and in section~\ref{sec:con} we will discuss proposed signatures of such scenarios, in the context of proposed signatures of baby-universe formation in the lab.

In the following section we shall outline inflation theory, which sets the current cosmological paradigm for the evolution of the early universe. It is the consideration of inflationary theory that led to one of the aforementioned recent proposals that cosmic black holes may house other universes~\cite{bib:2016JouCosAstPhy160060Gar}; and it was the development of inflation theory that led to the first proposals that a baby universe could be created in the laboratory~\cite{bib:1990NucPhy__B339417Fah}, sequestered within a micro black hole, in the 1980s~\cite{bib:1987PhyRev__D_35174Bla} (section~\ref{sec:buf}).

\section{\label{sec:inf}Inflation Theory as Precursor to Baby-Universe Formation}

The inflationary paradigm posits that the universe underwent a period of exponential expansion between approximately $10 ^{-36}$ and $10 ^{-32}$ seconds  after the big bang singularity. Below we outline the motivation for the theory, proposed in the 1980s, which also underpins the mechanism of baby-universe formation.

The fundamental idea behind inflation (and any contemporary cosmological model) is the recognition, from the general theory of relativity, that the universe is not a static entity, but has a complex and rich dynamics. This was cemented by the observations that led to the formulation of Hubble's law, according to which there is a clear linear correlation between our distance from other galaxies, and the velocity at which they are receding from us: the larger the distance, the larger, on average, the recession speed. It was then natural to imagine that, if other galaxies are now receding from us, in the past they should have been closer to us. Thus there could then be an event in the past, where the content of all galaxies in the universe could have been squeezed inside a very small region, where the expansion began. This special event in the history of the universe is now called the big bang singularity (or simply the big bang), and, following general relativity, it singles out not only the birth of our universe, but also of space and time themselves.

Another two crucial observations about our universe are the following:
\begin{enumerate}
	\item our universe looks about the same in every direction we look at: this is usually called \emph{isotropy};
	\item the universe appears to be the same around every point, i.e. no point seems to have special properties: this is usually called \emph{homogeneity}.
\end{enumerate}
They are the defining properties of the \emph{cosmological principle}. (Of course, if we look at the night sky, we can clearly discern differences between different regions of the sky, and the above two statements seem inconsistent with our visual experience. 
What they technically mean is that, by choosing a large enough scale~\footnote{Say, of the order of $100\,\mathrm{Mpc}$, where $1\,\mathrm{Mpc} \approx 3.1 \cdot 10 ^{19} \mathrm{km}$.}, the properties of the universe do satisfy homogeneity and isotropy when averaged over such a scale.) These properties of the universe are surprising because there is, in principle, no reason for places in the universe that are very far away from each other to share the same physical properties. On the other hand, as shown by the observations of the cosmic microwave background (CMB), anisotropies and inhomogeneities that we cannot account for are a very tiny effect. One possible explanation for the similarities we see across the sky is \emph{inflation}. According to inflation, the universe underwent a phase of strongly accelerated expansion almost immediately after the big bang. This expansion separated away regions that were in causal contact at the time, and, even after being separated, these regions kept the \emph{imprinting} of their (causally connected) early evolution. Inflation, then, resulted in the sudden creation of a very large volume of spacetime, as the universe grew from an initial tiny size, to something that then expanded into the vast collection of superstructures that we see today. Since inflation's development, numerous observations have found evidence corroborating predictions made by the paradigm~\cite{bib:2013PhyLet__B733112Gut}.

Inflation can be naturally related to the description of the fundamental interactions that comes from quantum field theory, and, in particular, with the physics of the quantum vacuum.

In field theory it is possible to consider field configurations that minimize the potential energy of the field. Since a field realizing one of these configurations has vanishing kinetic energy, then it is in a local minimum of its total energy.

Classically such a configuration is called an \emph{equilibrium configuration}: for small enough perturbations of an equilibrium configuration the field can acquire a small kinetic energy, but this will not be enough to move far away from the equilibrium configuration. In an ideal setup, the field will perform small oscillations around this configuration, but, in practice, some dissipative effects are also present. Then, the field will thus gradually lose the additional energy that it gained when the equilibrium state was perturbed, and settle back in the configuration corresponding to the local minimum of the energy. Thus, a local minimum of the potential energy is, not only, an equilibrium configuration, but a \emph{stable} one. In general, there can be many such configurations, possibly with different energies, and, classically, the system can realize in a stable way any such configurations.

\emph{Quantum mechanically}, however, the system has a non-zero probability to move to a lower energy minimum, even if a small perturbation does not provide it with enough energy to realize the same transition classically. This process is called \emph{quantum tunneling} (and it is responsible for many relevant physical phenomena, including the ones at the heart of the operation of modern electronic devices, like transistors). We are here interested in the effect of quantum tunneling on the vacuum structure~\cite{bib:1977PhyRev__D_15292Col,bib:1984NucPhy__B245481Rub,bib:1994PhyRev__D_40103Tan}, and, in particular, in the case when the vacuum structure interacts with gravity~\cite{bib:1980PhyRev__D_21330Col,bib:1994PhyRev__D_50644Tan}. Indeed, in curved spacetime, vacuum energy can act as a source of the gravitational field, and interesting effects may appear.

\begin{figure}
	\begin{center}
		\includegraphics[width=9cm]{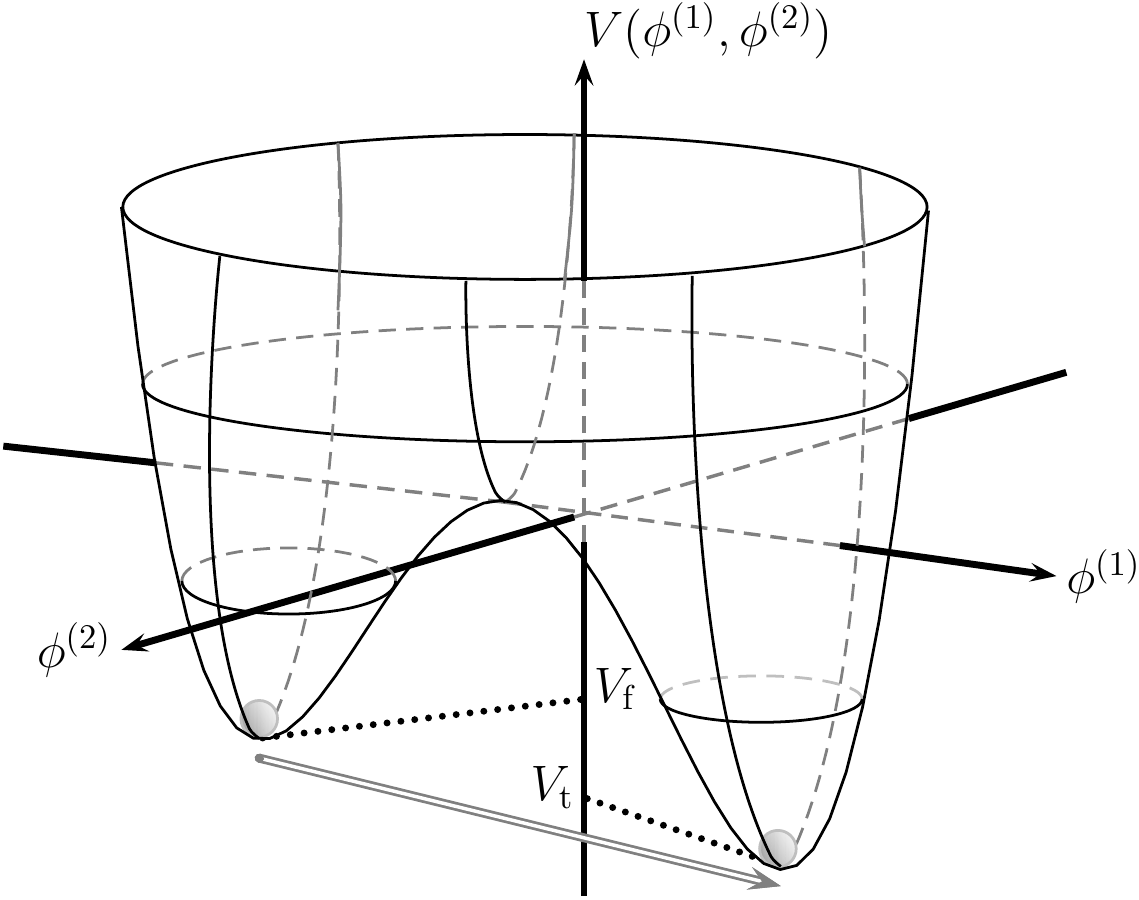}
		\caption{\label{fig:scafiepot}We show an example of a theory for two scalar fields, $\phi ^{(1)}$ and $\phi ^{(2)}$, with a potential possessing two local minima, $V _{\mathrm{f}}$ and $V _{\mathrm{t}}$ (with reference to the main text, here, $\vec{\phi} = (\phi ^{(1)} , \phi ^{(2)})$). The absolute minimum is $V _{\mathrm{t}}$, and corresponds to the \emph{true vacuum}, while $V _{\mathrm{f}}$ is the \emph{false vacuum}. If the field configuration corresponding to $V _{\mathrm{f}}$ is realized, the system is deep inside the potential well, and classically stable. However, quantum mechanically there is a non-zero probability for the system to tunnel from the false vacuum configuration to the true vacuum. We remember that, in general, the fields depend on time and position $\phi ^{(1,2)} = \phi ^{(1,2)} ( \vec{x} , t)$. Then, generically, there is no reason for this transition to affect, instantly, all space. It is more natural to expect a dynamical picture, in which the transition may take place at different times in different places: then, this would create different domains of the new vacuum state, evolving inside the old one.}
	\end{center}
\end{figure}
Following the work of Coleman and De Luccia, let us consider a model, in which the potential $V (\vec{\phi})$ for a collection of scalar fields $\vec{\phi}$ has the properties shown in figure~\ref{fig:scafiepot}. The contribution of the potential to the action, according to general relativity, is of the form
\[
	\int d ^{4} x \sqrt{-g} V (\vec{\phi})
	\ ,
\]
where, $\sqrt{-g}$ is the determinant of the spacetime metric, and, it is, in general, a non-trivial function of the coordinates. We then see that even a constant $V (\vec{\phi})$ contributes non-trivially, exactly because of the presence of $\sqrt{-g}$. More than this, if we write the action of general relativity in the presence of a cosmological constant
\[
	\int d ^{4} x \sqrt{-g} \{ R - 2 \Lambda \}
	,
\]
where $R$ is a function describing the curvature of spacetime and $\Lambda$ is a constant, we see that adding just a constant to $V (\vec{\phi})$ above, contributes exactly as a cosmological constant does in the last equation. This means that effects that could be considered as non-essential in the absence of gravity, can result in absolutely non-trivial physical effects in general relativity. One such consequence would have been to cause the early universe to inflate, expanding at an exponential rate. Another potential effect, the creation of a baby universe by triggering inflation
within a patch of our current universe, is discussed below.

\section{\label{sec:buf}Baby-Universe Formation}

\begin{figure}
	\begin{center}
		\includegraphics[width=11cm]{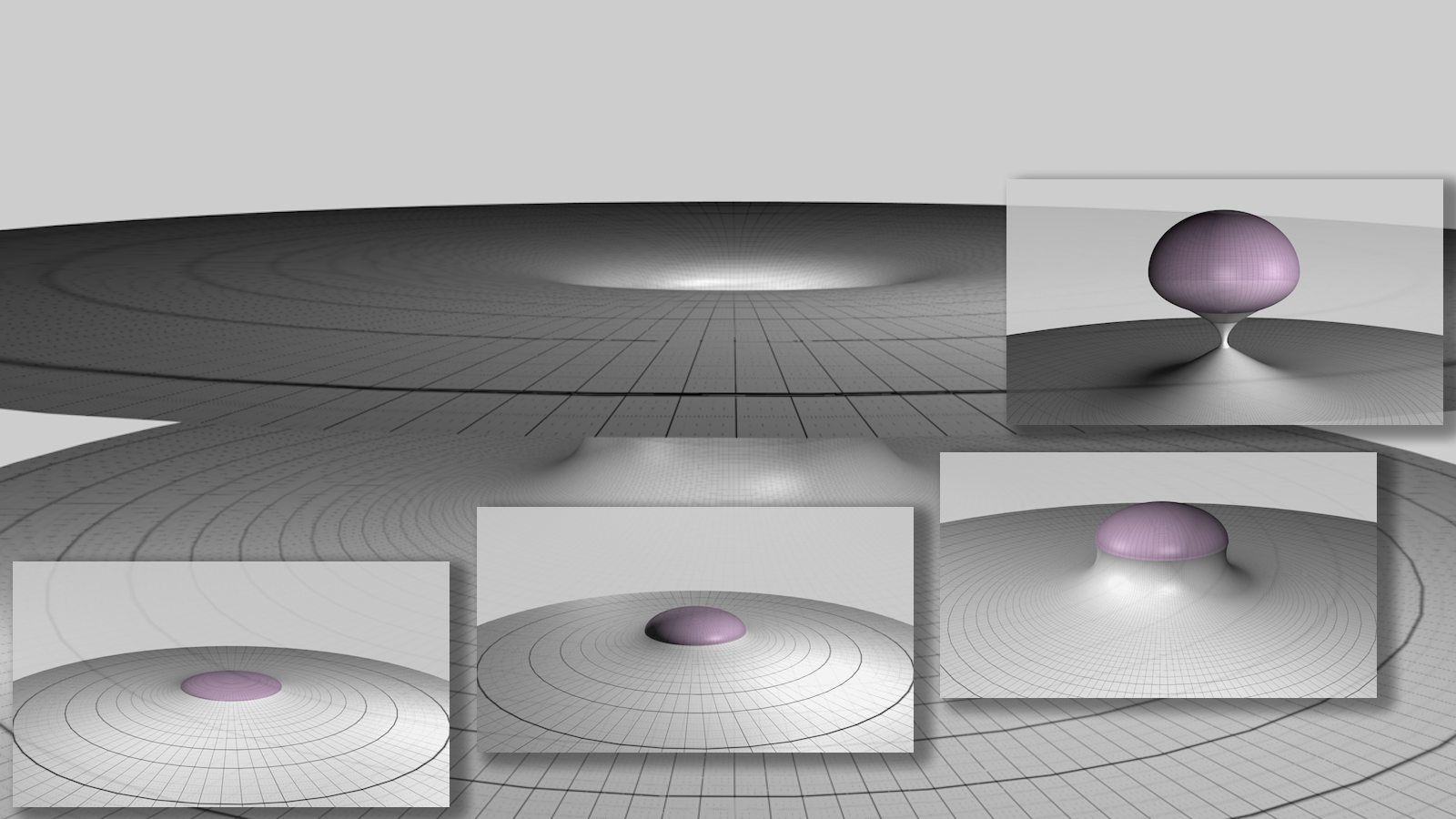}
		\caption{\label{fig:babunifor}The structure of wormhole spacetimes (visualized in the \emph{background} by the throat connecting two asymptotically flat regions) is a key element in the baby universe formation process. In the superimposed panels, from the bottom-left to the top-right, we show the initial vacuum region (\emph{first panel}), followed by the early time expansion before quantum tunneling (\emph{second panel}); then, we visualize the configuration immediately after the quantum tunneling process (\emph{third panel}, with the newly formed universe on the other side of a wormhole throat); finally, (\emph{fourth panel}) expansion continues with the baby universe creating its own space, and causally disconnecting from the parent spacetime.}
	\end{center}
\end{figure}

Baby-universe formation is a process that combines all the features discussed above; i.e., it is the result of the quantum mechanical transition between inequivalent vacuum states, when we take into account the coupling with gravity, and the characteristic features of black hole/wormhole spacetimes that we briefly discussed in section~\ref{sec:bhw}. All the above ingredients are essential for a consistent description of the process, which, qualitatively, can proceed as follows:
\begin{enumerate}
	\item[i.] In the current universe we consider a region, which does not realize the minimum energy (\emph{true}) vacuum.
	\item[ii.] Classically the region can be stable despite its raised energy, and is thus called a \emph{false} vacuum. 
	But quantum mechanically the probability that the false vacuum decays to the true vacuum by quantum tunneling is non-zero, and this process will occur, sooner or later. The interesting point is that quantum tunneling can be realized in two different ways:
	\begin{enumerate}
		\item[iii.a] the region tunnels in the same asymptotic region that the starting false vacuum region was; then, the region expands at the expense of the false vacuum region;
		\item[iii.b] the region tunnels into a different asymptotic region from the one inhabited by the starting false vacuum region; i.e., after the tunneling expansion takes place, the region that underwent tunneling evolves by creating its own spacetime, outside of the region from which it originated.
	\end{enumerate}
\end{enumerate}
In the first scenario (steps i., ii., and iii. a), the parent universe is cannibalized by the new, more stable vacuum; the susceptibility of our own universe's vacuum to such a catastrophic fate has been investigated in light of Higgs data from the Large Hadron Collider (LHC), see for instance \cite{bib:2017BigBanLitRooMer002}.
Baby-universe formation, however, is the distinct process described by i., ii., and iii.b: as a result, the region that underwent tunneling expands into a completely new universe (the baby universe). Initially, this child region is connected to its parent spacetime via a wormhole tunnel. From the perspective of the parent universe, this wormhole gateway appears only as a minute black hole. As the baby universe inflates, however, this wormhole tunnel is pinched closed, causally disconnecting the baby from the original, parent, spacetime~\cite{bib:1987PhyRev__D_35174Bla}.
This process can be intuitively related to our discussion of the causal structure of black hole spacetimes. Indeed, a standard tunneling process would correspond to a tunneling of the vacuum region that just makes it larger within the same parent spacetime. On the other hand, one can recognize that the tunneling process leading to the formation of a baby universe has the peculiar property to result in the production of an additional asymptotic region, i.e., tunneling also creates a new region of spacetime for the subsequent expansion of the baby universe outside of the parent spacetime~\cite{bib:2015JouExpThePhy120460}.

Note that the possibility for this baby-universe process to be realized strictly relies on the very peculiar properties of gravity described in terms of general relativity, i.e.
\begin{enumerate}
	\item[A.] there are spacetime structures that can contain several (causally disconnected) asymptotic regions;
	\item[B.] differences in vacuum energy can result in differences in the spacetime structure, because energy/momentum density and their flows act as sources of the gravitational field, according to the Einstein equations;
	\item[C.] spacetime can be (and is) created during cosmological evolution as the universe expands.
\end{enumerate}
Quantum tunneling plays two significant roles in the process: Firstly, it allows the transition between different vacuum states, described above, which would otherwise be impossible in a non-quantum scenario. Secondly, it also makes it possible for a baby universe to be created starting from 'reasonable' initial conditions~\cite{bib:1990NucPhy__B339417Fah}, as early recognized,
\begin{quote}
	``For inflation at a typical grand unified theory scale of $10 ^{14}\:\mathrm{GeV}$, for example, the universe emerges at $t \approx 10 ^{-35}\:\mathrm{s}$ from a region of false vacuum with a radius of only $\sim 10 ^{-24}\:\mathrm{cm}$ and a mass of only $\sim 10\:\mathrm{kg}$.''
\end{quote}
This result clearly raises the questions of how feasible it is to make a baby universe in a real-world laboratory -- potentially in a future particle accelerator -- and whether it may be possible to detect its production. Some proposals will be discussed in the next section, along with the notion that baby universes could arise spontaneously within our universe.

\section{\label{sec:con}Conclusion and Open Questions Regarding the Feasibility of Baby-Universe Production}

In this contribution, we have first considered black holes and wormholes in terms of the causal structure of spacetime. In particular, we have emphasized the richness and complexity of the nature of spacetime in general relativity. The consequences of this for the description of our universe can sometimes be fascinating and counterintuitive. In this context, the dynamics of spacetime naturally plays a central role, and we, qualitatively, reviewed such a role in the context of the physics of inflation,
and its connection with vacuum decay. A unique realization of all these ideas is baby-universe formation, a process in which a new domain of spacetime formed by quantum tunneling in a pre-existing universe can evolve by creating its own space, and eventually (causally) disconnecting from the parent spacetime. We remarked how compelling it is to consider that the initial conditions for the creation of such a universe are such that it is not unreasonable to expect that they could be realized somewhere within our universe: in the literature, this possibility is known as creating a universe in the laboratory.

When considering the possibility of making a baby universe in a laboratory within our universe, many issues need to be addressed. These include (but are not limited to):
\begin{enumerate}
	\item Whether it is possible to find, or generate, a seed (a patch of starting false vacuum), which can quantum tunnel to create a baby cosmos, as outlined in step iii.b, in the section above.
	\item Whether future (or even current) particle accelerators can reach the energies required to kick-start the inflationary process in that seed.
	\item Whether it would be possible to detect a baby universe -- should one be created -- by distinguishing it from an ``ordinary'' mini black hole.
	\item Whether baby universes are being spontaneously created in the vacuum -- and what physical implications this may have for theories of quantum gravity.
\end{enumerate}
Regarding the question of potential seeds for baby-universe formation, it has been proposed that inflation could be triggered in a monopole (a particle hypothesized to exist by some grand unified theories)~\cite{bib:1994PhyRev__D_50245Lin,bib:2006PhyRev__D_74024Sak}, if the monopole is subjected to high enough energies in a particle accelerator~\cite{bib:2017BigBanLitRooMer}, or in certain exotic matter that has negative mass~\cite{bib:1989PhyRevLet_63341Bar,bib:1990PhyRev__D_42262Har,bib:2008PhyRev__D_77125Gue} (which is hypothesized to exist in some versions of string theory). Such objects have yet to be found (although there are current searches underway to find monopoles at the LHC and elsewhere\footnote{Indeed, the Monopole and Exotics Detector at the LHC (MoEDAL) collaboration is searching for a particle carrying magnetic charge by using an array of plastic nuclear-track detectors~\cite{bib:2017PhyRevLet118061Ach}.}), which provides a significant obstacle to any attempts to carry out such a process in practice. It is also debatable whether next-generation particle
accelerators could reach energies required to generate a baby universe; this may require certain predictions of string theory to hold true, enhancing the effect of gravity at the scale of elementary particles~\cite{bib:2001PhyRevLet_87161Dim,bib:2002PhyRev__D_65056Gid,bib:2010Nat466426Mer}. At best, if these conditions hold, it has been claimed that planned accelerators, which may be built in coming decades, might reach the required energies~\cite{bib:2017BigBanLitRooMer001}.

Even assuming these hindrances are overcome, and a baby universe is generated in some future particle collider, it is not clear that physicists running the machine would be able to tell the difference between the production of a child cosmos and a standard micro black hole (predicted to exist by some string theory models~\cite{bib:2017BigBanLitRooMer001,bib:2001PhyRevLet_87161Dim,bib:2002PhyRev__D_65056Gid}). Current exotic physics searches at the LHC focus on the Hawking radiation signature (in the form of a shower of particles predicted to emanate in all directions from decaying micro black holes). It has been suggested that the rate of emission of this radiation would differ between a mini black hole that contains a baby universe and one that does not~\cite{bib:1991PhyRev__D_44333Lar}. Intriguingly, as mentioned in section~\ref{sec:bhw}, it has also recently been proposed that some inflationary models predict the existence of parallel universes hidden within supermassive black holes -- and that indirect observational evidence for their presence may be acquired from the mass distribution of black holes in the universe~\cite{bib:2017JouCosAstPhy170050Den}. In particular, it has recently been claimed that large mass, primordial black holes, might always contain a baby universe in their interior, because they would be formed by an expanding domain that cannot displace the exterior spacetime: in this case, a \emph{no-go} theorem that would put an upper-limit on the black hole mass, would be avoided as the process does not take place in an asymptotically flat spacetime, but in a cosmological background \cite{bib:2017JouCosAstPhy170050Den}.

Finally, the possibility that baby universes could be spontaneously generated in the vacuum of today's universe has been investigated. It has been claimed that the high energy density excitations,
responsible for ultraviolet divergences in quantum field theories, including quantum gravity, are likely to be the source of baby universes which carry them out of the original spacetime. Baby-universe formation could therefore be responsible for UV regularization in quantum field theories, which takes into account gravitational effects~\cite{bib:2008IntJouModPhy__D_17501Gue,bib:2008ProThePhy120985Ans}.

\section*{Acknowledgements}

E.~G. and Z.~M. gratefully acknowledge financial support from the Foundational Questions Institute (FQXi) through a mini grant, and would also like to thank the University of Bahamas for hosting the event ``A Big Bang in a Little Room''. Z.~M. would like to acknowledge partial support from the John Templeton Foundation for research on the topic. S.~A. would like to gratefully acknowledge hospitality by the Department of Physics of Kyoto University, while working on ideas related to the discussion in this paper.
Moreover, the authors would like to thank A.~Vilenkin, for insightful discussions about some of the ideas present in this paper.

\end{document}